# Systematic spatial and stoichiometric screening towards understanding the surface of ultrasmall oxygenated silicon nanocrystal

**Shanawer Niaz[1,2], Aristides D. Zdetsis[2], Emmanuel N. Koukaras[2], Oğuz Gülseren[1] and Imran Sadiq[3]**

[1]Department of Physics, Bilkent University, Ankara 06800, Turkey

[2]Molecular Engineering Laboratory, at the Department of Physics, University of Patras, Patras, GR-26500, Greece

[3]Centre of Excellence in Solid State Physics, University of the Punjab, Lahore, Pakistan

E-mail: shanawersi@gmail.com and shanawersi@fen.bilkent.edu.tr

**Abstract**

In most of the realistic *ab initio* and model calculations which have appeared on the emission of light from Si nanocrystals, the role of surface oxygen has been usually ignored, underestimated or completely ruled out. We investigate theoretically, by density functional theory (DFT/B3LYP) possible modes of oxygen bonding in hydrogen terminated silicon quantum dots using as a representative case of the $Si_{29}$ nanocrystal. We have considered Bridge-bonded oxygen (BBO), Doubly-bonded oxygen (DBO), hydroxyl (OH) and Mix of these oxidizing agents. Due to stoichiometry, all comparisons performed are unbiased with respect to composition whereas spatial distribution of oxygen species pointed out drastic change in electronic and cohesive characteristics of nanocrytals. From an overall perspective of this study, it is shown that bridge bonded oxygenated nanocrystals accompanied by Mix have higher binding energies and large electronic gap compared to nanocrystals with doubly bonded oxygen atoms. In addition, it is observed that the presence of OH along with BBO, DBO and mixed configurations further lowers electronic gaps and binding energies and trends. It is also demonstrated that oxidizing constituent besides their spatial distribution significantly alters binding energy and highest occupied molecular orbital (HOMO) and lowest unoccupied molecular orbital (LUMO) gap (HOMO-LUMO gap up to 1.48 eV) within same composition.

## 1. Introduction

Silicon nanocrystals (SiNCs) are very interesting nanomaterials whose potential is still not discovered completely or even understood in many respects. Compared to bulk silicon, the electronic properties of SiNCs are significantly dependent on their size. In general, these properties are extremely sensitive to the surface conditions of nanocrystals such as passivation, functionalization, spatial distribution of passivants, reconstruction etc. Silicon nanocrystals (SiNCs) possess quantum confinement effect, large ratios of surface area to volume, nontoxicity and biodegradability, leading to the use of SiNCs in a variety of fields such as microelectronics, optoelectronics, photovoltaics, in-vivo bioimaging, photosensitizing, drug delivery, lab-on-chip sensing, photocatalysis, phototherapeutics and much more [1-12]. Freestanding self-assembled SiNCs are often synthesized with surface hydrogen passivation which can be further oxidized [8]. But the role of surface oxygen has been usually ignored or underestimated, despite the evidence given by various experiments [13-15]. Many techniques have been used to explore the oxidation state of the Si atoms involved in bonding with surface oxygen [16-18]. In past years, a lot of efforts have been carried out in order to understand surface chemistry of silicon NCs due to the presence of oxidizing constituents both experimentally and theoretically [19-28]. For example, it was reported earlier [28] that for BBO containing NCs the red shift of band gap is found to be smaller compared to the DBO or complete hydroxylation. Furthermore, for hydroxyl passivation, energy gap is largely dependent on amount of OH on surface and their spatial distribution. Zdetsis et al. [25] demonstrated that BBO leads to more stable nanocrystal with large HOMO-LUMO gap and binding energies compared to DBO bonds with which we completely agree. Xiaodong et al. [26] performed similar oxygen treatment to hydrosylilated silicon NCs, where they conclude that BBO and OH hardly change the HOMO-LUMO gap at ground state (which is not true for hydrogen passivated NCs). Nazemi et al. [27] have investigated effect of spatial position and spatial distribution of BBO passivants on absorption spectra of hydrogen passivated NCs. According to their findings, spatial position can significantly effect on HOMO-LUMO gap, optical absorption and localization centers of frontier orbitals with which we agree. However, we have investigated that spatial distribution of DBO, Mix and OH also play vital role on surface chemistry in addition to BBO.

In this study, $Si_{29}$ nanocrystal (~1 nm) has been deliberately considered for electronic and cohesive investigations, which is not accidental as was demonstrated in our previous work [25]. Therefore, instead of random selection of oxygen bond formation and selective discussion on their characteristics, we present rather systematic density functional Theory (DFT) study. Hence oxidizing constituents of Bridge-bonded oxygen (BBO), Doubly-bonded oxygen (DBO), hydroxyl (OH) and mixed are examined with respect to their composition stoichiometry (identical isomers) and spatial position (distribution).

It is found that electronic and cohesive characteristics significantly change with (1) concentration, (2) spatial distribution and (3) type of oxygen bonding on the surface of silicon nanocrystals. In general, it is observed that bridge bonded oxygenated nanocrystals along with Mix have higher binding energies and large electronic gap compared to nanocrystals with doubly bonded oxygen atoms which is also true if hydroxyl group is also present. As far as concerned to the stability of NCs, we confirm from our results that BBO containing NCs are more stable than DBO whereas Mix bonding show intermediate behavior.

## 2. Model and approach

Initial structure of oxygen free hydrogen terminated $Si_{29}$ nanocrystal is selected from our previous studies [25, 29]. In this study, oxygenated $Si_{29}$ nanocrystals are constructed while taking special care of stoichiometry in order to present comprehensive and unbiased comparison. Fig. 1 represents some fully relaxed structures with respect to various modes of oxygen bonding and their concentrations. Detailed information about oxygen bond configurations along with spatial positions of those stoichiometrically identical isomers can also be seen in Table 1. As is mentioned above, we have considered various types of oxygen bonding, namely, Bridge-bonded oxygen (BBO or >O), Doubly-bonded oxygen (DBO or =O), hydroxyl (OH) and Mix. We put large emphasis on non-hydroxylated nanocrytals whereas selected cases of hydroxylated nanocrystals are also part of this research.

For example, in case of non-hydroxylated $Si_{29}O_6H_{24}$ nanocrystal, six BBO atoms are introduced in first subgroup $Si_{29}(>O)_6H_{24}$, six DBO atoms are introduced in second subgroup $Si_{29}(=O)_6H_{24}$ and in third subgroup a mixture of three BBO and three DBO atoms $Si_{29}(>O)_3(=O)_3H_{24}$ with exactly 50% ratio which keeps composition stoichiometry. Similar technique has been adopted for other non-hydroxylated nanocrystals where number of oxygen atom varies from $O_2$ to $O_{10}$ with even distribution. For hydroxylated nanocrystals, OH group is introduced in addition to the configuration adopted above. For example, in $Si_{29}O_{10}H_{24}$ nanocrystal, six BBO atoms are introduced in first subgroup $Si_{29}(>O)_6H_{20}(OH)_4$, six DBO atoms are introduced in second subgroup $Si_{29}(=O)_6H_{20}(OH)_4$ and in third subgroup a mixture of three BBO and three DBO atoms are introduced $Si_{29}(>O)_3(=O)_3H_{20}(OH)_4$ etc. In contrast with previous example of non-hydroxylated NCs, four hydrogen atoms have been replaced with four OH group. For comparison we have also included oxygen free hydrogen passivated $Si_{29}$ nanocrystal i.e. $Si_{29}H_{36}$ (Table 1).

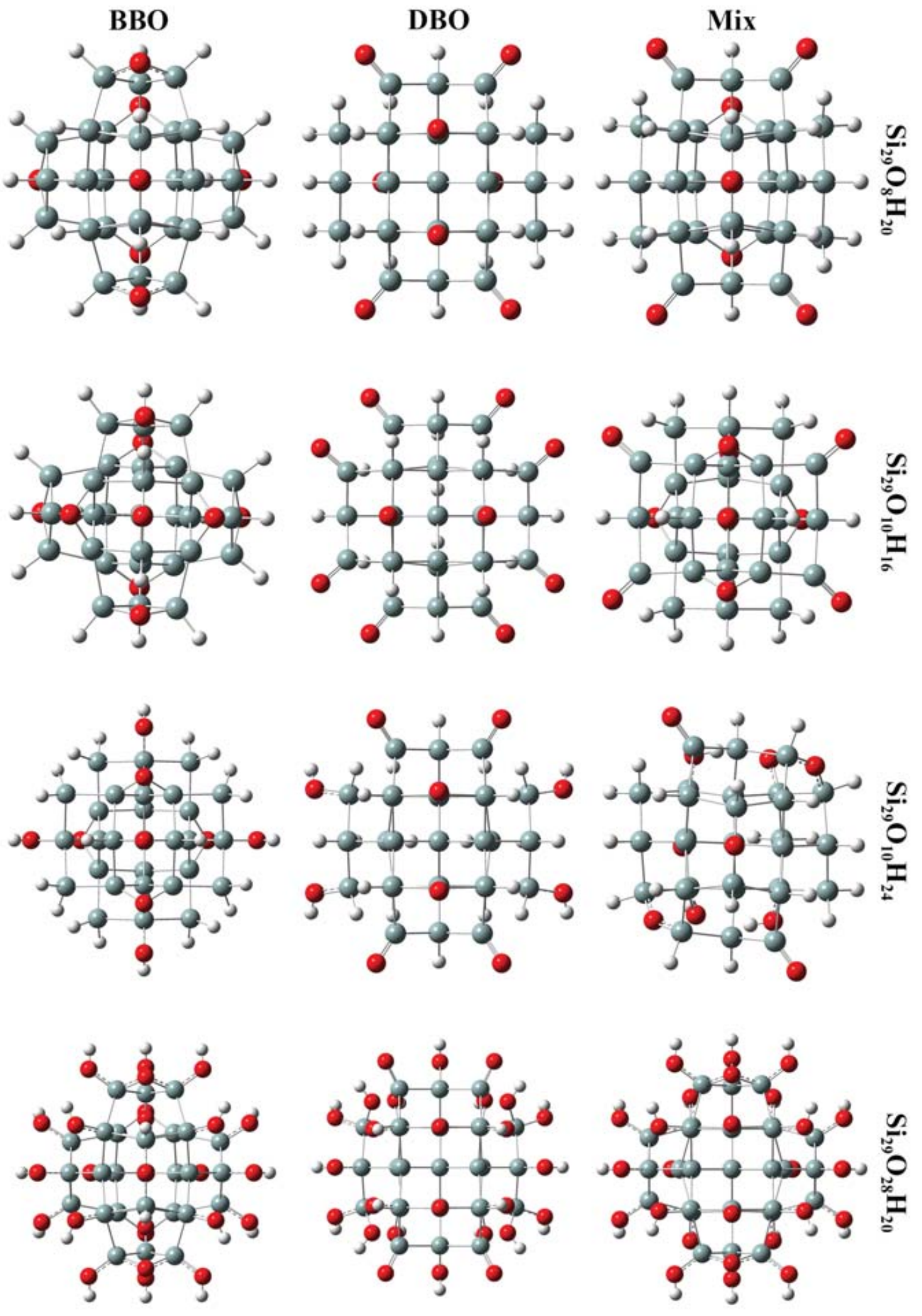

**Figure 1.** Optimized structures of non-hydroxylated $Si_{29}O_8H_{20}$, $Si_{29}O_{10}H_{16}$ and hydroxylated $Si_{29}O_{10}H_{24}$, $Si_{29}O_{28}H_{20}$ nanocrystals with BBO, DBO, OH and Mix configurations.

Spatial position or spatial distribution of oxidizing constituents, analogous to their stoichiometry, alters surface chemistry which cannot be neglected. Hence, structures are further divided into three categories depending upon the position of oxygen bonds while keeping their composition same. The choice of spatial position is strongly dependent on the availability of suitable silicon atom and their vacant neighborhood which can be saturate and avoid dangling bond. In this process no repetition is involved, hence, all possible combinations are included in this study (Table1). Due to the different configurations of nanocrystals depending on the number of oxygen atoms, their orientations and spatial distribution, symmetry of the structures differs compared to the symmetry of original structure i.e. $T_d$, which is obvious.

All calculations in this work are based on density functional theory (DFT) using the hybrid exchange-correlation functional of Becke, Lee, Yang and Parr (B3LYP) [30]. This functional has been shown to efficiently reproduce the band structure of a wide variety of materials, including c-Si, with no need for further numerical adjustments [31]. Convergence criteria for the SCF energies and for the electron density (rms of the density matrix), were placed at $10^{-7}$ au, whereas for the Cartesian gradients the convergence criterion was set at $10^{-4}$ au. Our calculations were performed with the TURBOMOLE [32] suite of programs using Gaussian atomic orbital basis sets SVP [4s3p1d] for Silicon and [3s2p1d] for Oxygen [33].

## 3. Results and discussion

### *3.1. Cohesive Properties:*

We have calculated binding/atomization energy of oxygenated silicon nanocrystal using following expression [25]:

$$BE_{NC} = N_{Si}E(Si) + N_O E(O) + N_H E(H) - E_{NC}[Si_{N_{Si}}O_{N_O}H_{N_H}]$$

where $E_{NC}[Si_{N_{Si}}O_{N_O}H_{N_H}]$ is total energy of nanocrystal, $E$(Si), $E$(O) and $E$(H) are the energies of silicon, oxygen and hydrogen atoms and $N_{Si}$, $N_O$, and $N_H$ are the number of silicon, oxygen and hydrogen atoms.

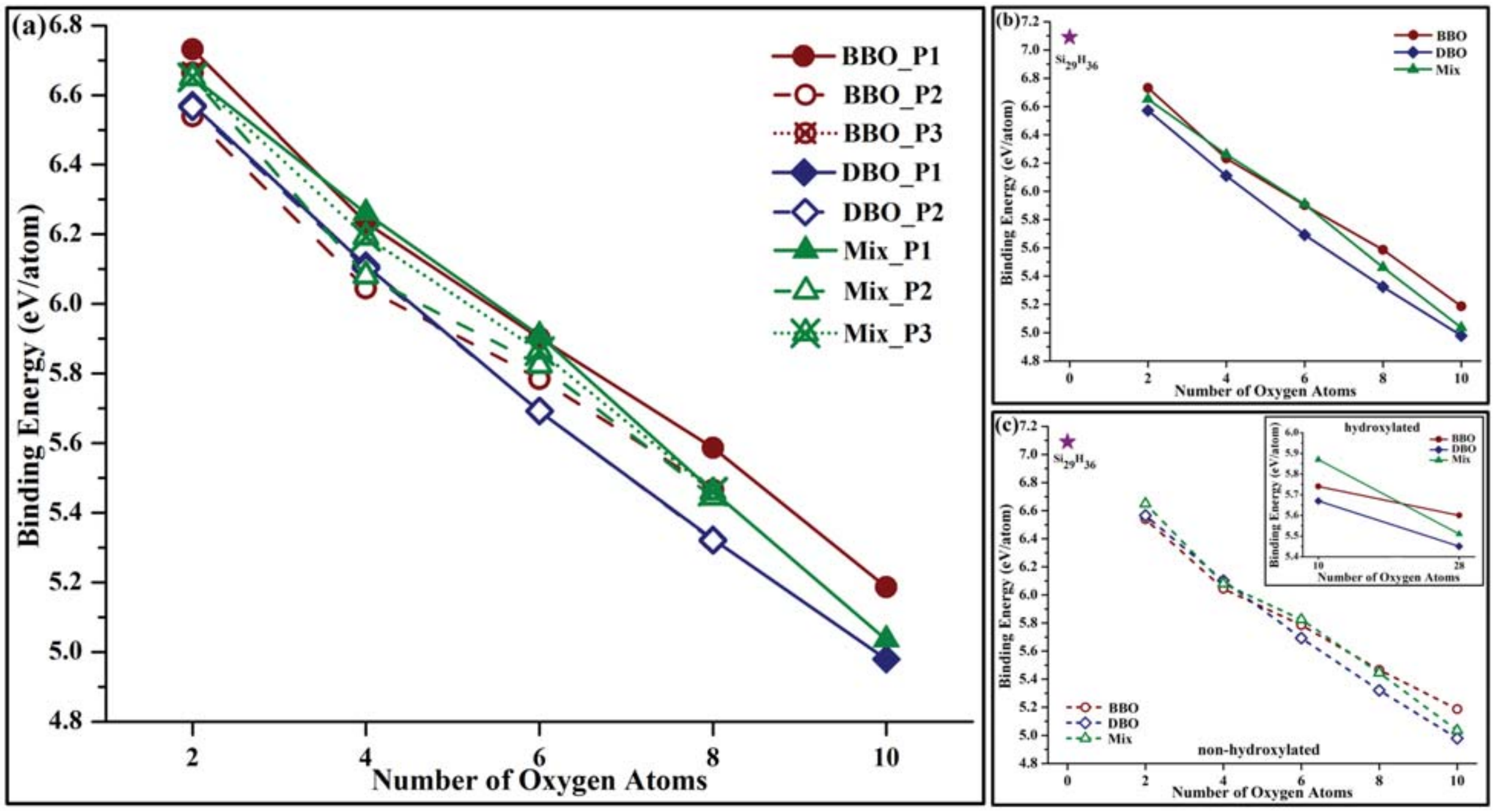

**Figure 2.** (a) Binding energy per heavy atom with respect to number of oxygen atoms diagram of all nanocrytals considered with possible modes of oxygen bonding, spatial positions and stoichiometric compositions. (b) and (c) represent compositions with highest and lowest binding energies values respectively whereas inset of (c) shows binding energies of selected hydroxylated nanocrytals. Magenta star shows binding energy of oxygen free silicon nanocrystal, $Si_{29}H_{36}$ for comparison.

Figure 2 shows binding energy per heavy atom with respect to the number of oxygen atoms on the surface of silicon nanocrystals. Further details about composition, binding energies and electronic gap etc for all NCs can be seen in Table1. In general, regardless of any interspecies comparison, the binding energy decreases with increasing number of oxygen atoms on the surface of silicon nanocrytals. As we can see in Figure 2(a) that BBO_P1 have highest binding energy values and DBO_P1 (and DBO_P2) have lowest binding energy values whereas all other configurations including mixed ones have intermediate values. Hence structural stability of BBO_P1 is not only higher compared with other spatial positions of BBO but also other form of oxygenation (DBO and Mix). It is interesting and rather clear from figure 2 that spatial position of DBO does not effect binding energy. However, in case of three different spatial positions of Mix oxygen configurations which contain both BBO and DBO, variation in binding energy is influenced mainly by the presence of BBO.

For clearer interpretation of binding energy comparision, we sorted out our results in further two groups. Hence, Figure 2(b) and 2(c) show nanocrystals with highest and lowest binding energy values respectively, extracted from Fig. 2a. The difference in binding energies for NCs with BBO and DBO is larger in 2(b) compared with 2(c). For example, taking into account for nanocrystals containing 8 oxygen atoms, binding energy per atom deference is 0.27 eV whereas in case of 2(c) it is 0.15 eV. To be more specific, an average interspecies binding energy difference is about 0.18 eV/atom, which cannot be neglected. It is also evident from figure that there is a competitive difference of binding energies between BBO and Mix which is again due to the presence of BBO in Mix nanocrystals (as is discussed above). It is worth mentioning that spatial positions of DBO do not much alter energies as shown in binding energy comparison of Fig. 2.

Inset diagram of figure 2(c) shows binding energy analysis for hydroxylated silicon nanocrystals including the presence of other oxidizing agents i.e. BBO, DBO and Mix. It is observed that the trend of cohesive characteristics do not much change but energy lowers compared to non-hydroxylated nanocrystals (see Fig 2a and 2c). Number of oxygen atoms shown in inset is 10 and 28 respectively which is sum of oxygen atoms from OH group and other oxidizing agents hence these energies may compare with the results of non-hydroxylated NCs having 6 and 8 oxygen atoms respectively.

### *3.2. Electronic Properties:*

We now focus on the electronic behavior of nanocrystals hence Figure 3 shows HOMO-LUMO gap according to the number of oxygen atoms for all possible combinations of NCs with respect to stoichiometry and their spatial positions. As far as the number of oxygen atoms is increasing the gap is decreasing significantly. All corresponding configurations of Figure 2 are included in Figure 3 in same sequence. Unlike the binding energies shown in figure 2, HOMO-LUMO gap energies are rather dispersed (between 2.33eV and 4.84eV) and show irregular behavior as number of oxygen atoms is increasing which is obvious due to the sensitive nature of silicon surface which can cause drastic impact on band gap. Again, one can observe a competition of gap energies between BBO_P1 and Mix_P1 configurations which contain large gap values compared to the DBO_P1 (and DBO_P2) and rest of the NCs have intermediate gap values.

Figure 3(b) and Figure 3(c) correspond to nanocrystals with largest and smallest HOMO-LUMO gap values, respectively, reproduced from Figure 3(a). One can observe that in Fig 3(b) the energy gap difference between BBO and DBO is 0.96 eV and in Fig 3(c) the difference is just 0.13 eV. However, we may draw an important analogy between number of oxygen atoms and interspecies HOMO-LUMO gap difference in such a way that the energy gap is uniformly decreasing as far as the NCs are fully saturated with oxygen. In other word, the gap difference is 1.12 eV in case of O=2 which is decreasing consistently to 0.27 eV for O=10 case.

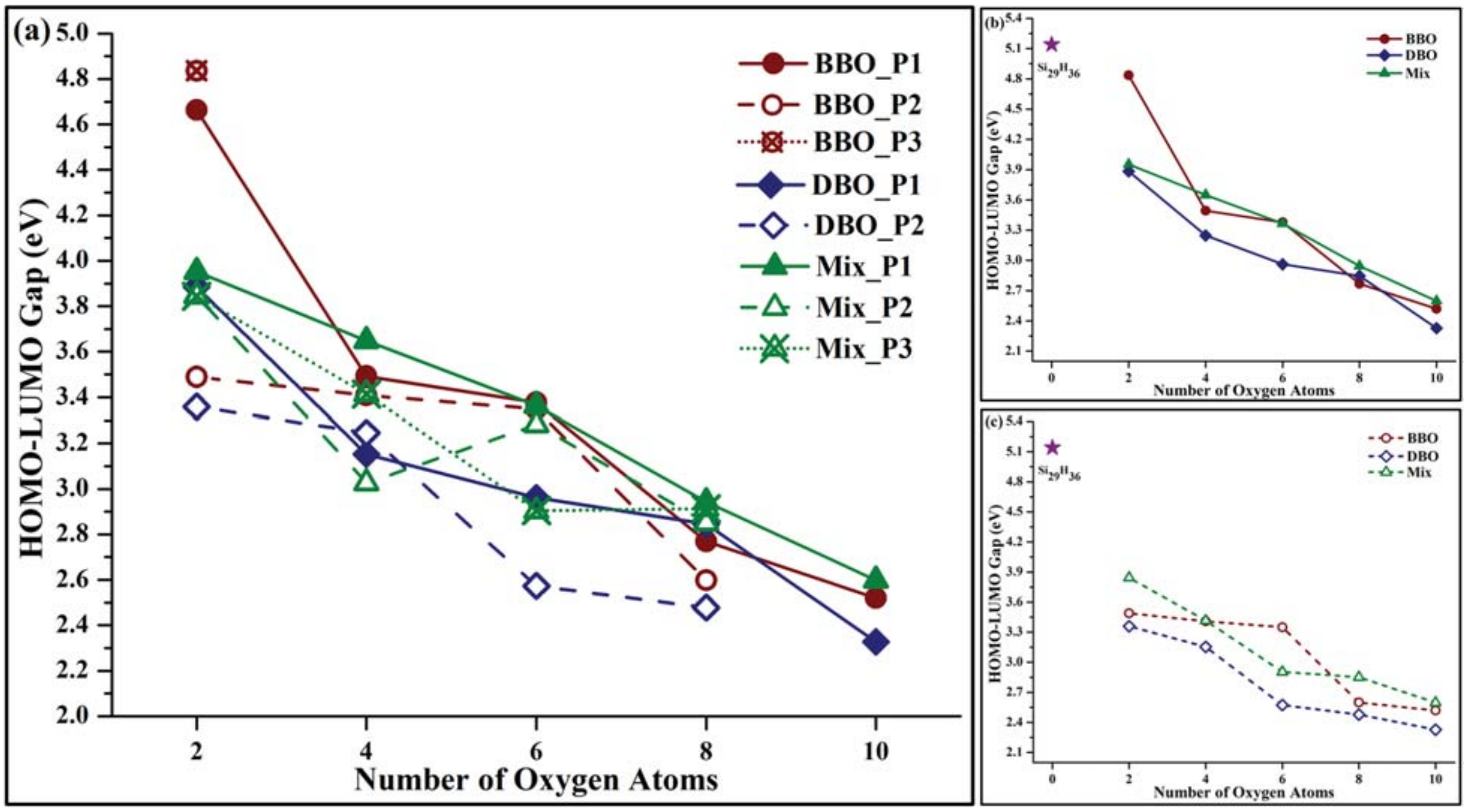

**Figure 3.** (a) HOMO-LUMO gap with respect to number of oxygen atoms of nanocrytals with possible modes of oxygen bonding, spatial positions and stoichiometric compositions. (b) and (c) represent nanocrystals with only largest and smallest HOMO-LUMO gap values respectively. Magenta star shows HL gap of oxygen free silicon nanocrystal, $Si_{29}H_{36}$ for comparison.

Once more, it is important to introduce contribution of spatial position (distribution) of oxidizing agents which can be understood while looking at binding energy difference and band gap difference of nanocrystals especially DBO containing isomers. For example, there is no difference in binding energies of DBO_P1 compared with DBO_P2 (Fig. 2a) whereas the gap difference is changed, dramatically, for identical composition (Fig. 3a). Hence our results for BBO support the demonstration of ref. [27] but this phenomenon can also be observed clearly when DBO or Mix oxidizing agents are present. Like hydroxylated nanocrystals along with other forms of oxygenation, non-hydroxylated nanocrystals exhibit similar trends (not shown here).

**Table 1:** Details of stoichiometrically identical isomers, oxidizing agents, binding energy per heavy atom and HOMO-LUMO gap for all configurations under study.

| Cluster | Formula | Modes of Bonds | B.E. (eV/atom) | H-L Gap (eV) | B.E. (eV/atom) | H-L Gap (eV) | B.E. (eV/atom) | H-L Gap (eV) |
|---|---|---|---|---|---|---|---|---|
| **Oxygen Free** | | | | | | | | |
| $Si_{29}H_{36}$ | $Si_{29}H_{36}$ | - | 7.09 | 5.14 | - | - | - | - |
| **Oxygenated (without OH group)** | | | | | | | | |
| | | | **Position-1** | | **Position-2** | | **Position-3** | |
| $Si_{29}O_2H_{32}$ | $Si_{29}(>O)_2H_{32}$ | BBO | 6.73 | 4.66 | 6.53 | 3.49 | 6.66 | 4.84 |
| | $Si_{29}(=O)_2H_{32}$ | DBO | 6.57 | 3.88 | 6.57 | 3.36 | - | - |
| | $Si_{29}(>O)_1(=O)_1H_{32}$ | Mix | 6.65 | 3.95 | 6.65 | 3.85 | 6.65 | 3.85 |
| $Si_{29}O_4H_{28}$ | $Si_{29}(>O)_4H_{28}$ | BBO | 6.23 | 3.49 | 6.04 | 3.41 | - | - |
| | $Si_{29}(=O)_4H_{28}$ | DBO | 6.11 | 3.15 | 6.10 | 3.25 | - | - |
| | $Si_{29}(>O)_2(=O)_2H_{28}$ | Mix | 6.26 | 3.65 | 6.08 | 3.03 | 6.19 | 3.42 |
| $Si_{29}O_6H_{24}$ | $Si_{29}(>O)_6H_{24}$ | BBO | 5.90 | 3.38 | 5.79 | 3.35 | - | - |
| | $Si_{29}(=O)_6H_{24}$ | DBO | 5.69 | 2.96 | 5.69 | 2.57 | - | - |
| | $Si_{29}(>O)_3(=O)_3H_{24}$ | Mix | 5.91 | 3.36 | 5.82 | 3.28 | 5.86 | 2.90 |
| $Si_{29}O_8H_{20}$ | $Si_{29}(>O)_8H_{20}$ | BBO | 5.59 | 2.77 | 5.47 | 2.60 | - | - |
| | $Si_{29}(=O)_8H_{20}$ | DBO | 5.32 | 2.84 | 5.32 | 2.48 | - | - |
| | $Si_{29}(>O)_4(=O)_4H_{20}$ | Mix | 5.46 | 2.94 | 5.44 | 2.85 | 5.46 | 2.91 |
| $Si_{29}O_{10}H_{16}$ | $Si_{29}(>O)_{10}H_{16}$ | BBO | 5.18 | 5.19 | - | - | - | - |
| | $Si_{29}(=O)_{10}H_{16}$ | DBO | 4.98 | 2.33 | - | - | - | - |
| | $Si_{29}(>O)_5(=O)_5H_{16}$ | Mix | 5.04 | 2.60 | - | - | - | - |
| **Oxygenated (with OH group)** | | | | | | | | |
| $Si_{29}O_{10}H_{24}$ | $Si_{29}(>O)_6H_{20}(OH)_4$ | BBO | 5.87 | 3.09 | - | - | - | - |
| | $Si_{29}(=O)_6H_{20}(OH)_4$ | DBO | 5.67 | 2.56 | - | - | - | - |
| | $Si_{29}(>O)_3(=O)_3H_{20}(OH)_4$ | Mix | 5.74 | 3.25 | - | - | - | - |
| $Si_{29}O_{28}H_{20}$ | $Si_{29}(>O)_8(OH)_{20}$ | BBO | 5.60 | 2.26 | - | - | - | - |
| | $Si_{29}(=O)_8(OH)_{20}$ | DBO | 5.45 | 2.11 | - | - | - | - |
| | $Si_{29}(>O)_4(=O)_4(OH)_{20}$ | Mix | 5.51 | 2.38 | - | - | - | - |

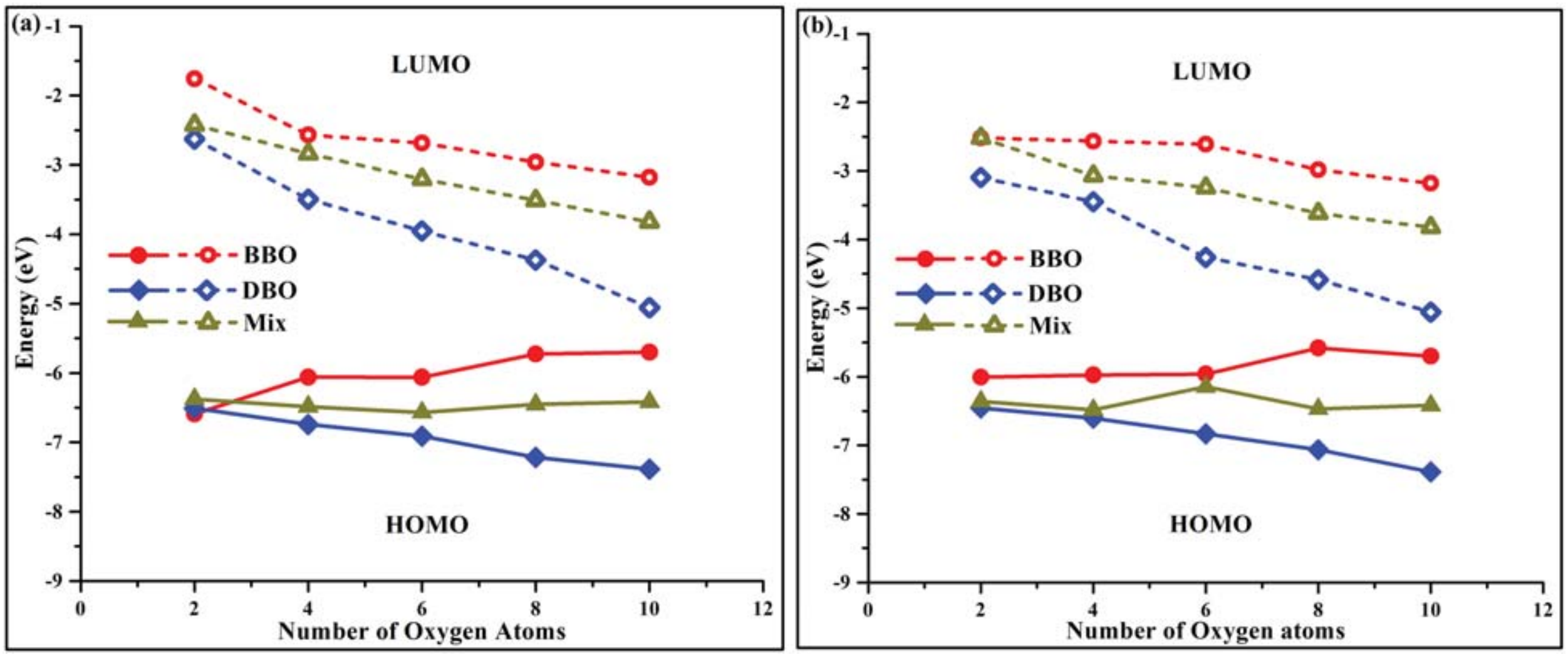

**Figure 4.** (a) and (b) show diagrams of HOMO and LUMO energies correspond to the gaps shown in Figure 2b and 2c respectively.

Figure 4 represents HOMO and LUMO energies with respect to increasing number of oxygen atoms for BBO, DBO and Mix configurations. HOMO and LUMO energies in Fig. 4a correspond to the nanocrystals with maximum energy values with respect to the spatial position of oxygen bonds. HOMO level is increasing for BBO and Mix configurations but decreasing in case of DBO whereas LUMO level is decreasing for all configurations in a uniform way as far as the number of oxygen numbers are increasing which is actually discrimination between the nature of DBO and other oxidizing constituents. The configurations of nanocrystals with minimum energies levels are shown in Figure 4b which represents similar tendency as shown in figure 4a for maximum gap energy values.

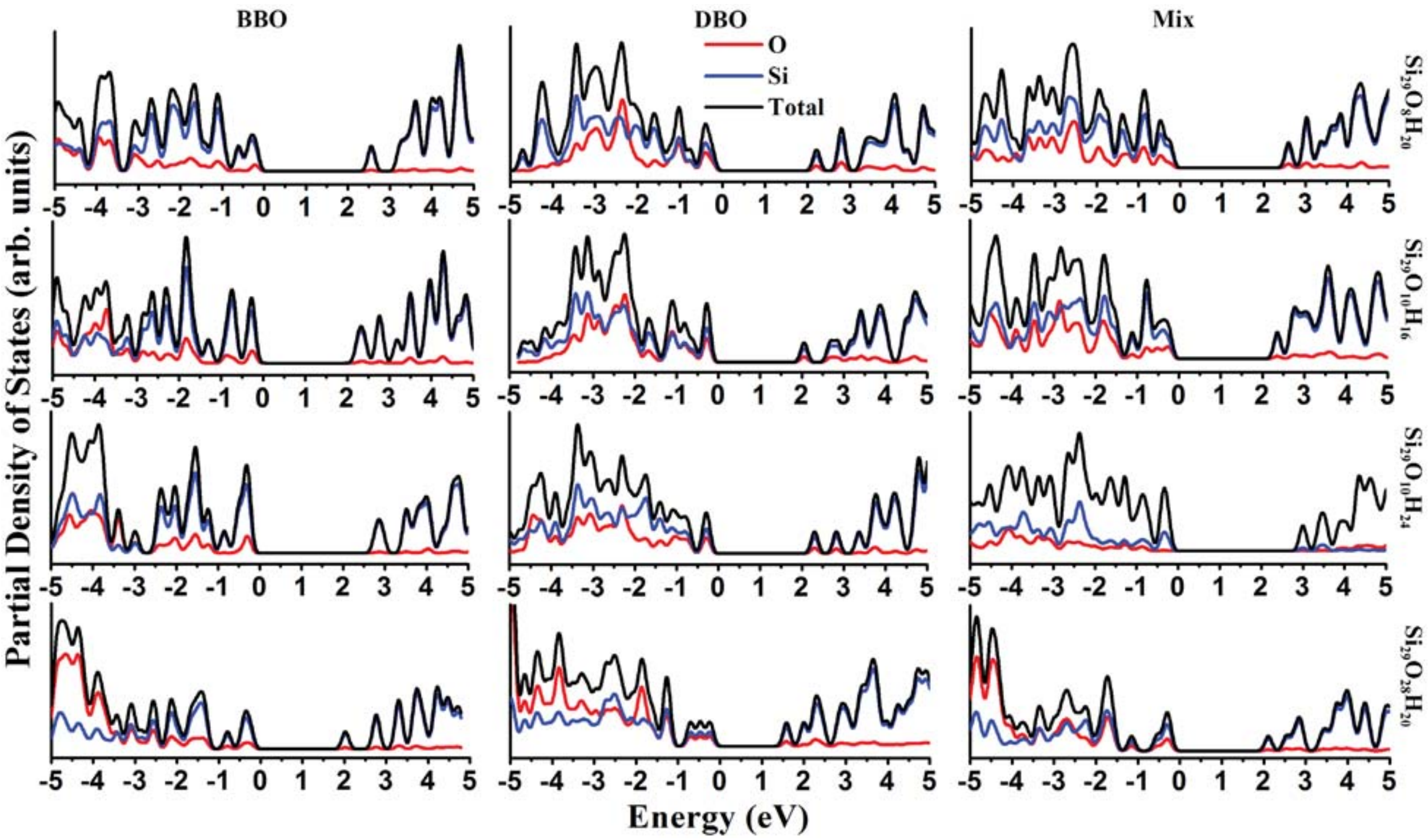

**Figure 5.** Partial density of states plots of non-hydroxylated $Si_{29}O_8H_{20}$, $Si_{29}O_{10}H_{16}$ and hydroxylated $Si_{29}O_{10}H_{24}$, $Si_{29}O_{28}H_{20}$ nanocrystals with BBO, DBO, and Mix configurations (geometries shown in figure 1). The Fermi level has been set to zero for clarity.

In addition, we represent Figure 5 which shows the partial density of states plot of selected non-hydroxylated ($Si_{29}O_8H_{20}$, $Si_{29}O_{10}H_{16}$) and hydroxylated ($Si_{29}O_{10}H_{24}$, $Si_{29}O_{28}H_{20}$) nanocrytals along with arbitrary position of BBO, DBO and Mixed configurations. Concerning BBO and Mixed configurations, the contribution of silicon atoms, near, HOMO is dominant compared to the oxygen atoms whereas in case of DBO both silicon and oxygen contribute almost equally. Also, one can easily notice from this figure that the LUMO level rapidly shift toward HOMO level when BBO is introduced which is not in case of DBO and less significant for Mix.

Finally, Figure 6 shows graphical representation of frontier molecular orbitals for two isomers of $Si_{29}O_2H_{32}$ nanocrystal with respect to spatial positions of BBO, DBO and Mix configurations. With the help of this representation, we demonstrated that spatial position significantly effect localization of frontier orbitals which is further investigated for various oxidizing agents. Hence in case of BBO of position-1, HOMO is localized only on left part of NC including one of the two oxygen atoms which alters the HOMO-1 where the distribution is on right side of the NC whereas LUMO is localized on entire NC along with both oxygen atoms and has overlap with UMOs. For DBO case, charge distribution is consistently identical in all considered frontier molecular orbitals and localization covers both oxygen atoms as well. In case of Mix configuration, there is complete overlap of HOMO and LUMO localization which is different compared with the rest of FMOs except HOMO-1.

Concerning position-2, for BBO configuration, HOMO and LUMO localized mainly at right end of the NC including both oxygen atoms. On the other hand, the OMOs and UMOs, localization differs from each pair and as of HOMO and LUMO where distribution on both oxygen atoms is not present at all. Like position-1, in DBO, the localization of HOMO and LUMO overlap where both doubly bonded oxygen atoms are included and for other frontier orbitals situation is not much different with varying concentration of the charge distributions. As far as concern to the Mix configuration, HOMO and LUMO localized strongly on bottom of the NC compared to other oxygen which has very less percentage of charge distribution. HOMO-1 localized mostly on the left including both oxygen atoms as can also be seen in case of HOMO-2 and UMOs.

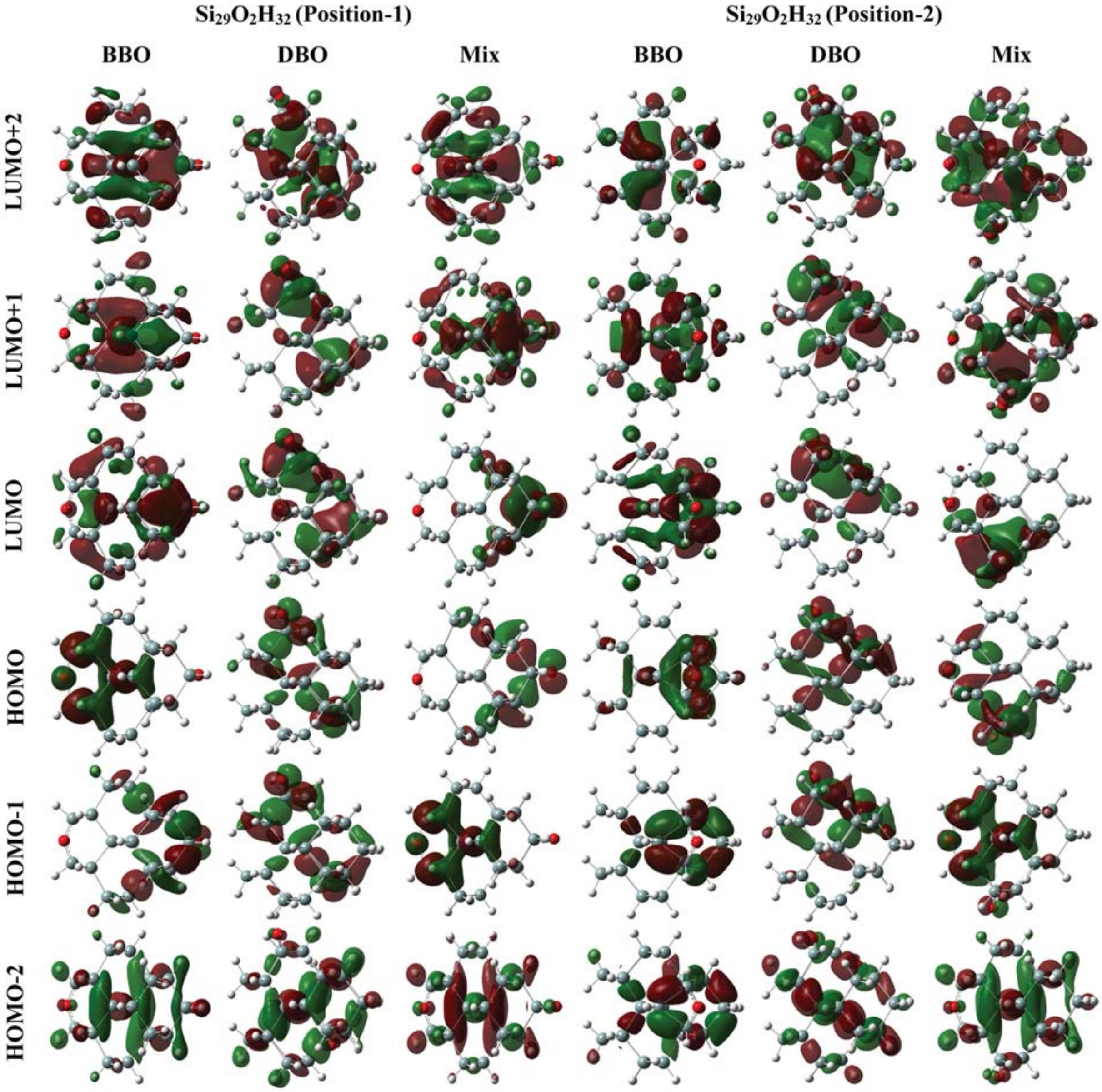

**Figure 6.** Graphical representation of various frontier molecular orbitals of $Si_{29}O_2H_{32}$ isomers.

## 4. Conclusions

We investigate theoretically, by density functional theory (DFT/B3LYP) possible modes of oxygen bonding such as Bridge-bonded oxygen (BBO), Doubly-bonded oxygen (DBO), hydroxyl (OH) and Mix of these oxidizing agents in hydrogen terminated silicon quantum dots using as a representative case of the $Si_{29}$ nanocrystal. Due to stoichiometry, all comparisons performed are unbiased with respect to composition whereas spatial distribution of oxygen species pointed out significant change in electronic and cohesive characteristics of nanocrytals. Bridge bonded oxygenated nanocrystals accompanied by Mix have higher binding energies and large electronic gap compared to nanocrystals with doubly bonded oxygen atoms. In addition, it is observed that the presence of hydroxyl group along with BBO, DBO and mixed configurations further lowers electronic gaps and binding energies and trends. It is also demonstrated that oxidizing constituent besides their spatial distribution substantially alters binding energy, HOMO-LUMO gap (up to 1.48 eV) and localization of frontier orbitals within same composition. For many scientific and industrial applications, a suitable range of gap energy can be achieved by appropriate selection/implementation of oxidizing agent and/or their spatial distribution which can be safe alternate of size dependent (or other) bandgap tunability.

**Acknowledgment**

This project is supported by The Scientific and Technological Research Council of Turkey (TÜBİTAK) under 2216 Research Fellowship Programme for International Researchers. Computational resources from Turkish Academic Network and Information Center (ULAKBİM) and National Centre for Physics Pakistan (NCP) are gratefully acknowledged.